\documentclass[12pt]{iopart}
\usepackage{graphicx}
\usepackage{bm}
\eqnobysec

\newcommand{\beq}{\begin{equation}}
\newcommand{\beqa}{\begin{eqnarray}}
\newcommand{\eeq}{\end{equation}}
\newcommand{\eeqa}{\end{eqnarray}}
\newcommand{\abs}[1]{\vert{#1}\vert}
\newcommand{\braket}[2]{\left<#1\mid#2\right>}
\renewcommand{\d}{{\rm d}}
\newcommand{\dow}{\downarrow}
\newcommand{\ds}{\displaystyle}
\renewcommand{\e}{{\rm e}}
\newcommand{\eps}{\varepsilon}
\newcommand{\frad}[2]{\displaystyle{\displaystyle#1\over\displaystyle#2}}
\newcommand{\ii}{{\rm i}}
\renewcommand{\k}{{(\rm K)}}
\newcommand{\mean}[1]{\langle#1\rangle}
\renewcommand{\phi}{\varphi}
\newcommand{\rr}{{\rm r}}
\newcommand{\si}{\sigma}
\renewcommand{\tr}{\mathop{\mathrm{tr}}\nolimits}
\newcommand{\up}{\uparrow}
\newcommand{\ve}[1]{{\bm#1}}
\newcommand{\E}{{\cal E}}
\newcommand{\D}{{\cal D}}
\renewcommand{\H}{{\cal H}}

\begin{document}

\title
[Tight-binding spectra on spherical graphs II:
the effect of spin-orbit interaction]
{Tight-binding electronic spectra on graphs with spherical topology II:
the effect of spin-orbit interaction}

\author{Y Avishai$^{1,2,3,4}$ and J M Luck$^3$}

\address{$^1$ Department of Physics and Ilse Katz Center for Nanotechnology,
Ben Gurion University, Beer Sheva 84105, Israel}

\address{$^2$ RTRA -- Triangle de la Physique, Les Algorithmes, 91190
Saint-Aubin, France}

\address{$^3$ Institut de Physique Th\'eorique\footnote{URA 2306 of CNRS},
CEA Saclay, 91191~Gif-sur-Yvette cedex, France}

\address{$^4$ Laboratoire de Physique des Solides\footnote{UMR 9502 of CNRS},
Universit\'e Paris-Sud, 91405 Orsay cedex, France}

\begin{abstract}
This is the second of two papers devoted to tight-binding electronic spectra
on graphs with the topology of the sphere.
We investigate the problem of an electron
subject to a spin-orbit interaction generated by the radial electric field
of a static point charge sitting at the center of the sphere.
The tight-binding Hamiltonian considered is a discretization
on polyhedral graphs of the familiar form ${\bm L}\cdot{\bm S}$
of the spin-orbit Hamiltonian.
It involves SU(2) hopping matrices of the form
$\exp({\rm i}\mu{\bm n}\cdot{\bm\sigma})$
living on the oriented links of the graph.
For a given structure, the dimensionless coupling constant $\mu$
is the only parameter of the model.
An analysis of the energy spectrum is carried out
for the five Platonic solids (tetrahedron, cube,
octahedron, dodecahedron and icosahedron) and the C$_{60}$ fullerene.
Except for the latter,
the $\mu$-dependence of all the energy levels is obtained analytically
in closed form.
Rather unexpectedly, the spectra are symmetric under the exchange
$\mu\leftrightarrow\Theta-\mu$,
where~$\Theta$ is the common arc length of the links.
For the symmetric point $\mu=\Theta/2$, the problem can be exactly mapped
onto a tight-binding model in the presence of the magnetic field
generated by a Dirac monopole, studied recently.
The dependence of the total energy at half filling
on $\mu$ is investigated in all examples.
\end{abstract}


\eads{\mailto{yshai@bgu.ac.il},\mailto{jean-marc.luck@cea.fr}}

\maketitle

\section{Introduction}
\label{intro}

The analogy between quantum dots and natural atoms is rather appealing,
and in many cases quantum dots are referred to
as artificial atoms (or molecules)~\cite{Imry}.
Within the physics of low-dimensional electronic systems, quantum dots and
natural molecules realize the ultimate extreme of zero dimension.
So far, most investigations have been
focused either on planar quantum dots or on quantum dots which
occupy a small volume (quantum box or cavity).
A novel class of zero-dimensional systems
which so far has not received much attention is realized
when electrons are confined to move on a compact surface of nanoscopic size.
The simplest class of such surfaces has the topology of the sphere.
An electron hopping between carbon atoms of a C$_{60}$ fullerene
(or its derivatives) provides the most natural candidate for such systems.

In the companion work~\cite{I} we have investigated in detail the spectra
of tight-binding electrons moving on polyhedral graphs with spherical topology,
subject to the radial magnetic field produced by a quantized magnetic charge.
This problem was solved for the five Platonic solids
(tetrahedron, cube, octahedron, dodecahedron, icosahedron), the C$_{60}$
fullerene and a couple of less symmetric objects (diamonds and prisms).
The main goal of the present work is to pursue this idea further by
including the electron spin and taking into account the spin-orbit interaction.
Within the same framework as in~\cite{I}, the single-particle energy
spectrum of an electron subject to a radial electric field
which generates a Rashba-type spin-orbit interaction~\cite{rashba}
is studied for the five platonic solids and the C$_{60}$ fullerene.

The spin-orbit interaction is known to have a profound impact in atomic,
nuclear and solid-state physics.
Confining our discussion to the latter field, a dramatic example of its
effect is the occurrence of an Anderson metal-insulator transition
in disordered two-dimensional electronic systems~\cite{loctwo}.
Recall that the spin-orbit interaction emerges as a natural consequence
of the Dirac equation,
when the low-energy sector is described by the Pauli equation,
and relativistic corrections are taken into account
by means of a systematic $1/c^2$ expansion~\cite{sodef}.
For an electron of mass $m$ and charge $-e$,
subject to an electrostatic potential $V(\ve r)$,
and therefore to a static electric field $\ve E(\ve r)=-\ve\nabla V(\ve r)$,
the spin-orbit term in the effective Hamiltonian is
\beq
\hat\H_{\rm SO}=\frac{e\hbar}{8m^2c^2}
\left(\ve p\cdot(\ve E\times\ve\si)+(\ve E\times\ve\si)\cdot\ve p\right),
\label{VSO}
\eeq
where $\ve p=-\ii\hbar\ve\nabla$ is the momentum operator
and $\ve\si$ is the vector of Pauli matrices,
so that $\ve S=\hbar\ve\si/2$ is the electron spin operator.
If the electrostatic potential $V(\bm r)=V(r)$ is {\it central},
the spin-orbit Hamiltonian~(\ref{VSO}) simplifies to
\beq
\hat\H_{\rm SO}=\frac{e}{2m^2c^2r}\,\frac{\d V(r)}{\d r}\,{\ve L\cdot\ve S},
\label{VSOcent}
\eeq
where $\ve L$ is the electron orbital angular momentum operator.
In particular, if the electron is confined to move on a spherical shell
with radius $R$,
the spin-orbit Hamiltonian~(\ref{VSOcent}) acquires the familiar form
\beq
\hat\H_{\rm SO}=C\ve L\cdot\ve S=\frac{C}{2}(\ve J^2-\ve L^2-\ve S^2),
\label{vLS}
\eeq
where $C$ is a constant and $\ve J=\ve L+\ve S$ is the total angular momentum.

The present work will be focused on the example
of the Coulomb potential produced by a static electric charge $q$
placed at the center of the sphere,
\beq
V(r)=\frac{q}{r},\quad\ve E(r)=\frac{q\ve r}{r^3}.
\label{vcou}
\eeq
For this potential, one has
\beq
C=-\frac{qe}{2m^2c^2R^3}.
\eeq
For the sake of completeness, we present at the end of Section~\ref{defs}
a discussion of the order of magnitude of the spin-orbit interaction,
although this can be found in many textbooks.

The Hamiltonian~(\ref{vLS}) has two eigenvalues $E_\pm$,
respectively corresponding to the vectors $\ve L$ and $\ve S$
being parallel and antiparallel.
If $\ell=0,1,2,\dots$ denotes the orbital quantum number,
the eigenvalues $E_\pm$ and their multiplicities $m_\pm$ read
\beqa
&&E_+=\frac{C\hbar^2\ell}{2},{\hskip 17.7truemm}m_+=2(\ell+1),\nonumber\\
&&E_-=-\frac{C\hbar^2(\ell+1)}{2},\quad m_-=2\ell.
\label{specLS}
\eeqa
This spectrum is not an even function of the coupling constant $C$,
except in the classical regime, i.e., in the $\ell\to\infty$ limit.
This lack of a symmetry is expected on physical grounds.
In the Coulomb case,
$C$ is indeed proportional to the product $qe$ of both charges.
Charges of the same sign ($C<0$) and charges of opposite signs $(C>0)$
indeed correspond to physically distinct situations,
which are not related by any symmetry.
Furthermore,
the situations where $\ve L$ and $\ve S$ are parallel and antiparallel
are also known to exhibit different features,
e.g.~in scattering theory~\cite{gw}.

Our main objective is to construct and study
natural discretizations of the spin-orbit Hamiltonian~(\ref{vLS}),
within a tight-binding model where the electron lives on the sites
(vertices) of a polyhedral graph drawn on the unit sphere
and executes nearest-neighbor hopping.
Our analysis will be based on an analogy with
the more conventional situation of tight-binding (spinless) electrons
subject to a given magnetic field $\ve B(\ve r)=\ve\nabla\times\ve A(\ve r)$.
In this case, the hopping of particles from site A to site B
is described by a hopping term of the form $a_{\rm A}^\dag U_{{\rm AB}}a_{\rm B}+\mathrm{h.c.}$
in the tight-binding Hamiltonian, where $U_{{\rm AB}}$ is a phase factor,
i.e., an element of the Abelian gauge group U(1).
It is generally accepted that the following expression,
known as the Peierls substitution~\cite{peierls}, is an appropriate choice:
\beq
U_{{\rm AB}}=\exp\left\{\frac{\ii e}{\hbar c}
\int_{\gamma({\rm A,B})}\ve A\cdot\d\ve r\right\},
\label{Peir}
\eeq
where $\gamma({\rm A,B})$ is a given continuous path joining site A to site B.
The phase factor so defined depends in general on the whole path $\gamma({\rm A,B})$,
and not only on the endpoints A and B
(see~\cite{gv} and~\cite{bbk} for recent investigations related to this matter).
For the present problem involving the non-Abelian gauge group SU(2),
the construction of the hopping terms requires some extra care.
Indeed, since SU(2) matrices do not commute among themselves,
an ordering prescription is needed in general.

The setup of the present paper is the following.
The model is introduced in Section~\ref{defs}.
The hopping matrices $U_{{\rm AB}}$,
which are elements of the non-Abelian SU(2) gauge group,
are evaluated for the two natural choices of shortest paths,
the straight line segment and the arc of a great circle.
A unique dimensionless parameter $\mu$ then appears in a natural way.
The main properties of the model, and especially its symmetries,
are studied in Section~\ref{props}.
In Section~\ref{spectra} the five regular polyhedra or Platonic solids
and the C$_{60}$ fullerene (modeled as a regular truncated icosahedron)
are investigated in detail.
The spectra of the tight-binding Hamiltonian
are respectively determined in Sections~\ref{spectetra} to~\ref{specfulle}.
The total energy at half filling is studied in Section~\ref{TotalE},
whereas Section~\ref{discussion} contains a short discussion.

\section{The model}
\label{Model}

\subsection{Definitions}
\label{defs}

In this work we consider a tight-binding model
defined on polyhedral graphs drawn on the unit sphere.
We denote by $V$ the number of vertices (sites),
by $L$ the number of links (bonds) and by $F$ the number of faces
of a polyhedron.
In the present case of spherical topology, the Euler relation reads
(see e.g.~\cite{wilson})
\beq
V-L+F=2.
\label{euler}
\eeq

In all the polyhedra considered in the following,
all the links have equal arc length~$\Theta$ (with $0<\Theta<\pi$).
For any pair of neighboring vertices A and B, we thus have
\beqa
&&\ve A\cdot\ve B=\cos\Theta,\quad
\ve A\times\ve B=\ve n_{\rm AB}\,\sin\Theta,
\label{abdef}
\eeqa
where $\ve A$ is the unit vector joining the center of the sphere to A,
and so on,
whereas $\ve n_{\rm AB}$ is the consistently oriented unit vector perpendicular
to $\ve A$ and $\ve B$, so as to have
\beq
\ve n_{\rm AB}=-\ve n_{\rm BA}.
\label{nn}
\eeq

The tight-binding model is defined by means of the Hamiltonian
\beq
\hat\H=\sum_{<{\rm AB}>}\left(\ve a^\dag_{\rm A}U_{{\rm AB}}\ve a_{\rm B}+\mathrm{h.c.}\right),
\label{ham}
\eeq
where the sum runs over the $L$ oriented links $<\!{\rm AB}\!>$
of the polyhedron, whereas
\beq
\ve a^\dag_{\rm A}=(a^\dag_{{\rm A}\up},a^\dag_{{\rm A}\dow}),\quad
\ve a_{\rm A}=\pmatrix{a_{{\rm A}\up}\cr a_{{\rm A}\dow}},
\eeq
where $a^\dag_{{\rm A}\sigma}$ and $a_{{\rm A}\sigma}$
are respectively the creation and annihilation operators
of an electron at site A with spin component $\sigma=\,\up$ or $\dow$,
and the matrices $U_{{\rm AB}}$
are elements of the non-Abelian gauge group SU(2),
i.e., $2\times2$ unitary matrices with unit determinant,
describing the spin-orbit coupling on an electron hopping from site A
to a neighboring site~B.

In analogy with the Abelian case described
by the Peierls substitution~(\ref{Peir}),
the SU(2) matrix $U_{\rm AB}$ is expressed as a path-ordered integral:
\beq
U_{\rm AB}={\rm P}\,\exp\left\{-\ii g\int_{\gamma({\rm A,B})}
(\ve E\times\ve\si)\cdot\d\ve r\right\},
\label{udef}
\eeq
where $\gamma({\rm A,B})$ is a given path joining site A to site B,
and $\ve E$ is the static electric field, as in~(\ref{VSO}).

The value of the coupling constant,
\beq
g=\frac{e}{4mc^2},
\label{gdef}
\eeq
is determined along the line of thought used in deriving
the Peierls substitution in the Abelian case~\cite{peierls},
and already considered e.g.~in~\cite{frostu} in the case of the SU(2) group.
The basic idea is to consider the spin-orbit term~(\ref{VSO})
as a perturbation of the free non-relativistic Hamiltonian
$\hat\H_0=\ve p^2/(2m)$ and to use the approximation
\beqa
\hat\H_0+\hat\H_{\rm SO}&=&
\frac{\ve p^2}{2m}+\frac{e\hbar}{8m^2c^2}
\left(\ve p\cdot(\ve E\times\ve\si)+\mathrm{h.c.}\right)\nonumber\\
&\approx&\frac{1}{2m}\Bigl(\ve p+\frac{e\hbar}{4mc^2}\ve E\times\ve\si\Bigr)^2
=-\frac{\hbar^2}{2m}\Bigl(\ve\nabla
+\ii\underbrace{\frac{e}{4mc^2}}_{\ds g}\ve E\times\ve\si\Bigr)^2.
\eeqa
The expressions inside parentheses in the second line
are respectively the SU(2) covariant momentum and derivative~\cite{frostu}.
It is worth noticing that the term obtained by expanding the square,
i.e., $g^2\hbar^2\ve E^2/(2m)$, if not neglected,
is a scalar potential which does not affect the spin physics anyhow.

The present work is restricted to the situation
where $\ve E$ is the electric field generated
by a static charge $q$ sitting in the center of the sphere,
given by~(\ref{vcou}), so that
\beq
(\ve E\times\ve\si)\cdot\d\ve r=-\frac{q}{r^3}(\ve r\times\d\ve r)\cdot\ve\si.
\eeq
Let us now make the hypothesis that the path $\gamma({\rm A,B})$
is planar, i.e., entirely contained in the OAB plane.
In this case, at every point of the path
the infinitesimal vector $\ve r\times\d\ve r$ is
perpendicular to the latter plane,
i.e., aligned with the vector $\ve n_{\rm AB}$
introduced in~(\ref{abdef}).
As a consequence, the path-ordering prescription is not needed,
and~(\ref{udef}) can be recast~as
\beq
U_{\rm AB}=\exp\left\{\ii gq\left(\int_{\gamma({\rm A,B})}
\frac{\ve r\times\d\ve r}{r^3}\right)\cdot\ve\si\right\}.
\label{utile}
\eeq

There are two natural choices for the path $\gamma({\rm A,B})$:

\noindent {\it Straight-line path.}
If $\gamma({\rm A,B})$ is the shortest path in three-dimensional space,
i.e., the straight line segment joining the points A and B,
$\ve r$ can be parametrized as
\beq
\ve r=(1-t)\ve A+t\ve B\quad(0\le t\le1).
\eeq
We have then
\beqa
&&\ve r\times\d\ve r=\sin\Theta\,\ve n_{\rm AB}\,\d t,\quad
r=(1-4\sin^2(\Theta/2)\,t(1-t))^{1/2},
\eeqa
so that~(\ref{utile}) yields
\beq
U_{\rm AB}=\exp(\ii gq\sin\Theta\,I(\Theta)\,\ve n_{\rm AB}\cdot\ve\si),
\eeq
with
\beq
I(\Theta)=\int_0^1\frac{\d t}{(1-4\sin^2(\Theta/2)\,t(1-t))^{3/2}}
=\frac{1}{\cos^2(\Theta/2)},
\eeq
i.e.,
\beq
U_{\rm AB}=\exp(2\ii gq\tan(\Theta/2)\,\ve n_{\rm AB}\cdot\ve\si).
\label{usdef}
\eeq

\noindent {\it Great-circle path.}
If $\gamma({\rm A,B})$ is the shortest path on the sphere,
i.e., the arc of the great circle passing through A and B,
$\ve r$ can be parametrized as
\beq
\ve r=\frac{\sin(\Theta-\tau)\,\ve A+\sin\tau\,\ve B}{\sin\Theta}
\quad(0\le\tau\le\Theta).
\eeq
We have $r=1$, whereas
\beq
\ve r\times\d\ve r=\ve n_{\rm AB}\,\d\tau,
\eeq
so that~(\ref{utile}) yields
\beq
U_{\rm AB}=\exp(\ii gq\Theta\,\ve n_{\rm AB}\cdot\ve\si).
\label{ugdef}
\eeq

For both choices of the path $\gamma({\rm A,B})$,
Equations~(\ref{usdef}) and~(\ref{ugdef}) yield
the same expression for the SU(2) matrix $U_{\rm AB}$:
\beq
U_{\rm AB}
=\exp(i\mu\,\ve n_{\rm AB}\cdot\ve\si)
=\cos\mu+\ii\sin\mu\,\ve n_{\rm AB}\cdot\ve\si,
\label{u}
\eeq
which gives the desired discretization of the familiar spin-orbit operator
$\ve L\cdot\ve S$ recalled in~(\ref{vLS}).
The fact that the hopping matrix $U_{\rm AB}$ involves a vector
parallel to $\ve A\times\ve B$
was already noticed in the case of a Rashba spin-orbit interaction
in semiconductors~\cite{sr,eagkv}.

For a given polyhedron, the model therefore has one single parameter,
the dimensionless coupling constant $\mu$.
Re-inserting for a while the physical radius $R$ of the sphere,
and using the expression~(\ref{gdef}) of the coupling constant $g$,
we are left with the following expression for $\mu$,
for both choices of the path $\gamma({\rm A,B})$:
\beq
\mu=\eps\times\left\{\matrix{
2\tan(\Theta/2)\hfill&\hbox{(straight-line path),}\hfill\cr
\Theta\hfill&\hbox{(great-circle path),}\hfill
}\right.
\label{mudef}
\eeq
with
\beq
\eps=\frac{gq}{R}=-\frac{mR^2C}{2}=\frac{qe}{4mc^2R}.
\label{epsdef}
\eeq

The dimensionless (positive or negative) number $\eps$
gives a measure of the strength of the spin-orbit interaction.
Its expression~(\ref{epsdef})
can be made more transparent by introducing the (positive or negative)
atomic number~$Z$, such that the charge at the center of the sphere is $q=Ze$.
One has then
\beq
\eps=\frac{Z\alpha^2}{4}\,\frac{a_0}{R},
\eeq
where $\alpha=e^2/(\hbar c)\approx1/137$ is the fine structure constant
and $a_0=\hbar^2/(me^2)$ is the Bohr radius,
whereas $R$ is the radius of the spherical sample.
Although the number $\eps$ is a priori very small,
due to the factor $\alpha^2\sim10^{-4}$,
it is allowed to become appreciable in the following two ways.
First, there is a priori no upper limit on the value of $Z$,
as the charge $q=Ze$ is treated in this work as a static classical charge.
Second, spin-orbit interactions can be many orders of magnitude
larger in solid materials than in vacuum,
due to Bloch electrons moving close to atomic nuclei
with relativistic velocities~\cite{Engel}.

The dependence of the parameter $\mu$ on the angle $\Theta$
in~(\ref{mudef}) also deserves a word of comment.
The same linear growth at small angles,
i.e., $\mu\approx\eps\Theta$, holds for both paths,
in accord with the expectation that we are dealing with a bona fide
discretization of the familiar spin-orbit Hamiltonian~(\ref{vLS}).
On the contrary, the regime of large angles ($\Theta\to\pi$)
exhibits two very different kinds of behavior:
$\mu$ remains finite in this limit in the case of a great-circle path,
whereas it diverges in the case of a straight-line path,
as the latter passes very near the center of the sphere,
where the electric field becomes infinitely large.

Throughout the following,
we shall adopt the theoretical viewpoint
of considering~$\mu$ as an arbitrary parameter,
forgetting both about its physical origin
and about its expression~(\ref{mudef}).
Of course, the arc length $\Theta$ of the links is bound to keep
its value, dictated by the geometry of the graph under consideration.
The dependence of energy spectra on $\mu$ will be investigated systematically,
starting with a study of its symmetries in the next section.

\subsection{Properties}
\label{props}

We now turn to a discussion of various properties
of the tight-binding Hamiltonian $\hat\H$ of the problem,
defined in~(\ref{ham}),
where the SU(2) matrices $U_{{\rm AB}}$ are given by~(\ref{u}),
putting a special emphasis onto symmetries.

\smallskip\noindent{\it Hermitian matrix representation.}
The relation~(\ref{nn}) ensures that the matrices $U_{{\rm AB}}$ obey
\beq
U_{{\rm AB}}=U_{\rm BA}^{-1}=U_{\rm BA}^\dag.
\label{uustar}
\eeq
The Hamiltonian $\hat\H$ is therefore represented by a $2V\times2V$
Hermitian matrix $\H$, whose rows and columns are labeled
by a couple $({\rm A}\sigma)$ where ${\rm A}=1,\dots,V$ denotes a site
and $\sigma=\,\up$ or $\dow$ is a spin index, such that
\beq
\H_{({\rm A}\sigma)({\rm B}\tau)}=\left(U_{{\rm AB}}\right)_{\sigma\tau}.
\label{hu}
\eeq
The equation for the energy eigenvalues $E_a$,
labeled by the integer $a=1,\dots,2V$,
and the corresponding eigenfunctions $\ve\psi_{{\rm A},a}$ reads
\beq
E_a\ve\psi_{{\rm A},a}=\sum_{{\rm B}({\rm A})}U_{{\rm AB}}\ve\psi_{{\rm B},a},
\label{eigen}
\eeq
where ${\rm B}({\rm A})$ runs over the neighbors of ${\rm A}$.
More explicitly,
\beq
E_a\psi_{{\rm A}\sigma,a}=\sum_{{\rm B}({\rm A})}\sum_{\tau=\up,\dow}
\left(U_{{\rm AB}}\right)_{\sigma\tau}\psi_{{\rm B}\tau,a}.
\eeq

\smallskip\noindent{\it Sum rules.}
The spectrum of the Hamiltonian $\hat\H$ obeys the following sum rules
\beq
\sum_aE_a=0,\quad\sum_aE_a^2=4L.
\label{sumrules}
\eeq
where the sums run over the $2V$ eigenvalues $E_a$,
repeated according to their mul\-ti\-pli\-ci\-ties.
The first sum equals $\tr\H=\sum_{\rm A}\tr U_{\rm AA}=0$.
This sum rule is a common feature to all tight-binding Hamiltonians
with only non-diagonal matrix elements.
The second sum equals $\tr\H^2=\sum_{{\rm AB}}\tr(U_{\rm AB}U_{\rm BA})=4L$.
Equation~(\ref{uustar}) indeed implies that each link $<\!{\rm AB}\!>$
gives two contributions
equal to $\tr(U_{\rm AB}U_{\rm AB}^\dag)=\tr\ve1=2$,
i.e., the number of spin degrees of freedom.

\smallskip\noindent{\it Kramers degeneracy.}
All the energy levels of the Ha\-m\-il\-tonian $\hat\H$
are at least twofold degenerate, because of time-reversal symmetry,
embodied in the Kramers theorem~\cite{kramers,LL}:
if $\ve\psi_{{\rm A},a}$ is a solution of~(\ref{eigen}),
another independent solution of the same equation,
with the same energy $E_a$,
is provided by the spinor $\ve\psi^\k_{{\rm A},a}$, where
\beq
\ve\psi^\k=\ii\si_{\rm y}\ve\psi^\star,\quad\hbox{i.e.,}\quad\left\{
\matrix{\psi^\k_\up=-\psi_\dow^\star,\cr\psi^\k_\dow=\psi_\up^\star.\hfill}
\right.
\label{trs}
\eeq
Here and throughout the following, the star denotes complex conjugation.

\smallskip\noindent{\it Homogeneous modes on regular polyhedra.}
In the case of the five regular polyhedra,
one can predict the existence of a
twofold degenerate eigenvalue associated with homogeneous modes.
Using the expression~(\ref{u}) of the matrices $U_{\rm AB}$,
the eigenvalue equation~(\ref{eigen}) can be recast~as
\beq
E\ve\psi_{\rm A}=\sum_{\rm B(A)}
(\cos\mu+\ii\sin\mu\,\ve n_{\rm AB}\cdot\ve\si)\ve\psi_{\rm B}.
\label{eigenab}
\eeq
For each site A, consider the vector
\beq
\ve W_{\rm A}=\ve A\times\sum_{\rm B(A)}\ve B
=\sin\Theta\sum_{\rm B(A)}\ve n_{\rm AB}.
\eeq
In the case of a regular polyhedron, one has $\ve W_{\rm A}=\ve 0$ by symmetry.
Indeed, $\ve W_{\rm A}$ is perpendicular to $\ve A$,
and in the plane perpendicular to $\ve A$
it has the $p$-fold rotational symmetry of the polyhedron,
where $p$ is the coordination number of the vertices.
$\ve W_{\rm A}$ therefore clearly vanishes.
As a consequence,~(\ref{eigenab}) shows that
the homogeneous wavefunction $\ve\psi_{\rm A}=\ve\chi$,
where $\ve\chi$ is a constant spinor,
independent of the site A, is an eigenfunction of the Hamiltonian $\hat\H$.
The corresponding twofold degenerate energy is $E=p\cos\mu$.

\smallskip\noindent{\it Semi-periodicity.}
The matrices $U_{{\rm AB}}$ given in~(\ref{u})
obey $U_{\rm AB}(\mu+\pi)=-U_{\rm AB}(\mu)$.
The energy eigenvalues therefore obey the same property,
referred to as {\it semi-periodicity}:
they are changed into their opposites if~$\mu$ is changed to $\mu+\pi$.
It is also worth noticing
that the spectrum of the Hamiltonian $\hat\H$ is
not an even function of~$\mu$,
in spite of the identity $U_{\rm AB}(-\mu)=U_{\rm AB}^\dag(\mu)$.
This lack of symmetry has already been emphasized
in the simpler example of the spin-orbit Hamiltonian~(\ref{vLS}).

\smallskip\noindent{\it $\mu\leftrightarrow\Theta-\mu$ symmetry.}
The Hamiltonian $\hat\H$ has the following less obvious symmetry.
For each site A, consider the spin operator in the direction of $\ve A$,
\beq
S_{\rm A}=\ve A\cdot\ve\si.
\eeq
One has clearly $S_{\rm A}^2=1$.
Furthermore, using the identity
\beq
(\ve a\cdot\ve\si)(\ve b\cdot\ve\si)
=\ve a\cdot\ve b+\ii(\ve a\times\ve b)\cdot\ve\si,
\label{abid}
\eeq
one can check that
\beqa
U_{\rm AB}(\Theta)=S_{\rm A}S_{\rm B}
\label{sasb}
\eeqa
for any pair of neighboring vertices A and B,
where $U_{\rm AB}(\Theta)$ is a shorthand for
the matrix $U_{\rm AB}$ given in~(\ref{u}) for $\mu=\Theta$.
Some algebra involving a repeated use of the same identity~(\ref{abid})
allows one to prove the more general relation
\beq
U_{\rm AB}(\Theta-\mu)=S_{\rm A}U_{\rm AB}(\mu)S_{\rm B}.
\label{simil}
\eeq
As a consequence, the Hamiltonians $\hat\H(\mu)$ and $\hat\H(\Theta-\mu)$
have the same spectrum.
More precisely, if $\ve\psi_{\rm A}$ is an eigenfunction
of $\hat\H(\mu)$ with energy $E$,~(\ref{simil}) shows that
$\ve\chi_{\rm A}=S_A\ve\psi_{\rm A}$ is an eigenfunction
of $\hat\H(\Theta-\mu)$ with the same energy $E$.

\smallskip\noindent{\it Special values of $\mu$.}
The following two values of the parameter $\mu$:
\beq
\mu_0=\Theta/2,\quad\mu_1=\Theta/2+\pi,
\label{spec}
\eeq
are special in several respects.
The $\mu\leftrightarrow\Theta-\mu$ symmetry
and the semi-periodicity imply that $\mu_0$ and $\mu_1$
are symmetry axes of the energy spectrum, if displayed as a function of~$\mu$.
The Kramers degeneracy and the $\mu\leftrightarrow\Theta-\mu$ symmetry
imply that all energy levels are at least fourfold degenerate,
i.e., that all the multiplicities are multiples of 4.

There is also a striking correspondence between the present problem
at the special value $\mu_0$
and the tight-binding problem in the presence of a magnetic monopole
investigated in~\cite{I}.
The mapping between both problems goes as follows.
For $\mu=\mu_0=\Theta/2$ one has
\beq
2\cos\mu_0\,U_{\rm AB}(\mu_0)=1+S_{\rm A}S_{\rm B}.
\label{onesasb}
\eeq
This formula, which is essentially equivalent to~(\ref{sasb}),
suggests to introduce
the local basis of eigenstates of the spin operators $S_{\rm A}$.
Denoting by $(\theta_{\rm A},\phi_{\rm A})$ the spherical coordinates
of A, such that
\beq
\ve A=(\sin\theta_{\rm A}\cos\phi_{\rm A},\sin\theta_{\rm A}\sin\phi_{\rm A},
\cos\theta_{\rm A}),
\eeq
the spinors $\ve\chi_{\rm A}^\pm$ such that
$S_{\rm A}\ve\chi_{\rm A}^\pm=\pm\ve\chi_{\rm A}^\pm$ read
\beq
\ve\chi_{\rm A}^+=\pmatrix{
\cos\frad{\theta_{\rm A}}{2}\cr
\sin\frad{\theta_{\rm A}}{2}\,\e^{\ii\phi_{\rm A}}},\quad
\ve\chi_{\rm A}^-=\pmatrix{
-\sin\frad{\theta_{\rm A}}{2}\,\e^{-\ii\phi_{\rm A}}\cr
\cos\frad{\theta_{\rm A}}{2}}.
\eeq
The spinors $\ve\chi_{\rm A}^\pm$ are changed into one another
by time-reversal symmetry, according to~(\ref{trs}).
Expanding the eigenfunctions of $\hat\H$ as
\beq
\ve\psi_{\rm A}=u_{\rm A}\ve\chi_{\rm A}^++v_{\rm A}\ve\chi_{\rm A}^-,
\label{uv}
\eeq
some algebra using~(\ref{onesasb}) shows that the amplitudes
$u_{\rm A}$ and $v_{\rm A}$ obey the following two
scalar tight-binding equations:
\beq
Eu_{\rm A}=\sum_{\rm B(A)}t_{\rm AB}u_{\rm B},\quad
Ev_{\rm A}=\sum_{\rm B(A)}t^\star_{\rm AB}v_{\rm B},
\label{euev}
\eeq
where the hopping rate $t_{\rm AB}$ is given by
\beq
2\cos\mu_0\,t_{\rm AB}
=\braket{\ve\chi_{\rm A}^+}{\ve\chi_{\rm B}^+}
=\cos\frac{\theta_{\rm A}}{2}\cos\frac{\theta_{\rm B}}{2}
+\sin\frac{\theta_{\rm A}}{2}\sin\frac{\theta_{\rm B}}{2}
\,\e^{\ii(\phi_{\rm B}-\phi_{\rm A})}.
\label{tabdef}
\eeq
The expression~(\ref{tabdef}) of $t_{\rm AB}$ can be
drastically simplified as follows.
Using trigonometric identities and the relation
\beq
\cos\Theta
=\cos\theta_{\rm A}\cos\theta_{\rm B}
+\sin\theta_{\rm A}\sin\theta_{\rm B}\cos(\phi_{\rm B}-\phi_{\rm A}),
\eeq
the expression for the square modulus of $t_{\rm AB}$
can be shown to boil down to $\abs{t_{\rm AB}}^2=1$.
The hopping rates are therefore phase factors.
Setting $t_{\rm AB}=\exp(\ii\omega_{\rm AB})$, we obtain
\beq
\cos\omega_{\rm AB}
=\frac{\cos\Theta+\cos\theta_{\rm A}+\cos\theta_{\rm B}+1}
{4\cos\frad{\Theta}{2}\cos\frad{\theta_{\rm A}}{2}\cos\frad{\theta_{\rm B}}{2}}.
\eeq
This expression can be recognized as one of the variants
of the spherical Heron formula giving the solid angle
of a spherical triangle in terms of its arc lengths~\cite{kk,edm},
recalled in the Appendix of~\cite{I}.
We thus obtain
\beq
t_{\rm AB}=\exp\left(\frac{\ii\Omega_{\rm NAB}}{2}\right),
\eeq
where $\Omega_{\rm NAB}$ is the solid angle of the oriented spherical triangle
NAB, where N is the North pole of the unit sphere.
It can be checked that phases and orientations are consistent,
so that the product of phase factors living on the anticlockwise
oriented links around any face equals $\exp(\ii\Omega/2)$,
where $\Omega$ is the spherical angle of the face under consideration.
This is precisely the requirement to describe the magnetic flux
generated by a magnetic monopole of unit charge ($n=1$)
sitting at the center of the sphere.

We have therefore shown that the spectrum of the present problem
at the special value $\mu=\mu_0=\Theta/2$
consists of two independent copies of the spectrum
of the magnetic monopole problem for $n=\pm1$.
One may wonder how the Hamiltonian $\cal\H$,
which is invariant under time reversal,
can have at the special point $\mu=\mu_0=\Theta/2$
the same spectrum as the Hamiltonian of the magnetic monopole problem,
for which the time-reversal symmetry is broken.
The key to the answer is that the expansion~(\ref{uv}),
which is the step that seemingly breaks time-reversal symmetry,
amounts to performing a unitary transformation $\D$ which
brings the $2V\times 2V$ Hamiltonian matrix $\H$ into the block diagonal form
\beq
\overline\H=\D\H\D^\dag=\pmatrix{\cal\H_+&0\cr0&\cal\H_-},
\eeq
where $\cal\H_\pm$ are the $V\times V$ Hamiltonian matrices
of the magnetic monopole problem with respective magnetic charges $n=\pm1$.
The full Hamiltonian $\overline\H$ is invariant under time reversal.
The operators $\cal\H_+$ and $\cal\H_-=\cal\H_+^\dag$
separately break time-reversal symmetry,
but they are changed into one another by time reversal.
Each of them brings one copy of the spectrum of the magnetic monopole problem.

\section{Polyhedra and their spectra}
\label{spectra}

In this section we investigate the spectrum
of the tight-binding Hamiltonian $\hat\H$ for the five regular polyhedra
or Platonic solids
and for the fullerene, modeled as a symmetric truncated icosahedron.
In the case of the Platonic solids,
all the properties derived in Section~\ref{props} will be checked
against our analytic expressions of the energy spectra.
The fullerene will be special in the following two respects.
Its energy spectrum will not be obtained analytically,
albeit from the numerical diagonalization of an explicit $120\times120$ matrix.
The homogeneous modes described in Section~\ref{props} are absent,
as the fullerene is not sufficiently symmetric to allow them.

All these polyhedra have been described in detail in~\cite{I}.
Table~\ref{tabpoly} lists a few of their geometrical characteristics
which will be useful in the following.

\begin{table}[!ht]
\begin{center}
\begin{tabular}{|l|c|c|c|c|c|c|}
\hline
polyhedron&$V$&$L$&$F$&$p$&$q$&$\cos\Theta$\\
\hline
tetrahedron&4&6&4&3&3&$-1/3$\\
cube&8&12&6&3&4&$1/3$\\
octahedron&6&12&8&4&3&0\\
dodecahedron&20&30&12&3&5&$\sqrt{5}/3$\\
icosahedron&12&30&20&5&3&$\sqrt{5}/5$\\
fullerene&60&90&32&3&$\left\{\matrix{5\cr6}\right.
$&$(80+9\sqrt{5})/109$\\
\hline
\end{tabular}
\end{center}
\caption{Geometrical characteristics of the polyhedra considered in this work:
numbers $V$ of vertices, $L$ of links, $F$ of faces,
coordination number (number of neighbors of a vertex)~$p$,
number of sides of a face $q$, expression of $\cos\Theta$,
where the arc length $\Theta$ of the links
has been introduced in~(\ref{abdef}).}
\label{tabpoly}
\end{table}

\subsection{The tetrahedron}
\label{spectetra}

The tetrahedron is the simplest of the Platonic solids.
It consists of 4 trivalent vertices,~6 links and 4 triangular faces.

\begin{figure}[!ht]
\begin{center}
\includegraphics[angle=90,width=.3\textwidth]{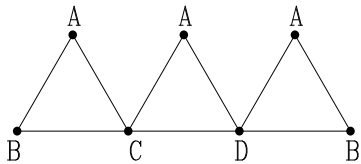}
\caption{
\label{vtetra}
Planar representation of the tetrahedron.}
\end{center}
\end{figure}

Throughout the following it will be advantageous to unwrap
the polyhedra around an axis of high symmetry, say of order $r$,
to be used as the $z$-axis~\cite{sm}.
For all the Platonic solids,
the order $r$ of rotational symmetry can be chosen to be the larger of
the integers~$p$ and $q$.
The planar representation of the tetrahedron thus obtained,
emphasizing the vertices and the links between them,
is shown in Figure~\ref{vtetra}.
Some vertices and links may have several occurrences,
to be identified by the inverse procedure of wrapping
the planar representation onto the sphere.
This planar representation is an efficient tool
to find the Cartesian coordinates of the vertices,
making an optimal use of symmetries.
Vertices at the same height on the plot have the same $z$ coordinate,
whereas their coordinates in the $xy$-plane are obtained from each other
by rotations by the commensurate angles $2\pi k/r$ for $k=1,\dots,r$.
Table~\ref{ctetra} lists the coordinates of the vertices thus obtained.

\begin{table}[!ht]
\begin{center}
\begin{tabular}{|c|c|c|c|}
\hline
vertex&$x$&$y$&$z$\\
\hline
A&0&0&1\\
\hline
B&$2\sqrt{2}/3$&0&$-1/3$\\
C&$-\sqrt{2}/3$&$\sqrt{6}/3$&$-1/3$\\
D&$-\sqrt{2}/3$&$-\sqrt{6}/3$&$-1/3$\\
\hline
\end{tabular}
\end{center}
\caption{Cartesian coordinates of the vertices of the tetrahedron.
Horizontal lines separate groups of vertices having the same $z$ coordinate.}
\label{ctetra}
\end{table}

\begin{figure}[!ht]
\begin{center}
\includegraphics[angle=90,width=.4\textwidth]{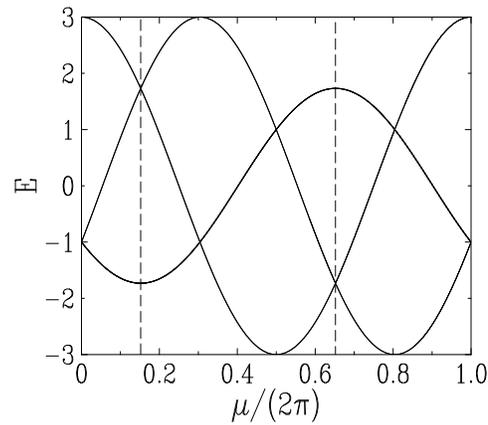}
\caption{
\label{sptetra}
Plot of the energy spectrum of the tetrahedron
against $\mu/(2\pi)$ over one period.
Vertical dashed lines:
symmetry axes at the special values of $\mu$ given in~(\ref{spec}).}
\end{center}
\end{figure}

The $8\times8$ Hamiltonian matrix $\H$ has been 
constructed from these coordinates by using~(\ref{u}) and~(\ref{hu}).
This matrix is too complex to be diagonalizable by hand.
For each of the five Platonic solids
this task has been performed with the help of the software MACSYMA.
The energy eigenvalues and their multiplicities
are listed in Table~\ref{tabtetra}.
Horizontal lines separate groups of levels related to each other
by the $\mu\leftrightarrow\Theta-\mu$ symmetry.
The levels $E_1(\mu)$ and $E_2(\mu)$ are interchanged
by this symmetry, whereas $E_3(\mu)$ is symmetric by itself.
One has indeed
\beq
E_2(\mu)=E_1(\Theta-\mu),\quad
E_3(\mu)=-\sqrt{3}\,\cos(\mu-\Theta/2).
\eeq
The energy spectrum is shown in Figure~\ref{sptetra}
as a function of $\mu/(2\pi)$ over one period.
The vertical dashed lines show the symmetry axes of the spectrum
at the special values of $\mu$ given in~(\ref{spec}).

\begin{table}[!ht]
\begin{center}
\begin{tabular}{|c|c|c|}
\hline
$a$&$E_a(\mu)$&$m_a$\\
\hline
1&$3\cos\mu$&2\\
2&$-\cos\mu+2\sqrt{2}\,\sin\mu$&2\\
\hline
3&$-\cos\mu-\sqrt{2}\,\sin\mu$&4\\
\hline
\end{tabular}
\end{center}
\caption{Energy levels $E_a(\mu)$ of the tetrahedron
and their multiplicities $m_a$.
Horizontal lines separate groups of levels related to each other
by the $\mu\leftrightarrow\Theta-\mu$ symmetry.}
\label{tabtetra}
\end{table}

\subsection{The cube}
\label{speccube}

\begin{figure}[!ht]
\begin{center}
\includegraphics[angle=90,width=.3\textwidth]{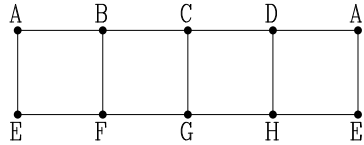}
\caption{
\label{vcube}
Planar representation of the cube.}
\end{center}
\end{figure}

The planar representation of the cube is shown in Figure~\ref{vcube}.
Table~\ref{ccube} lists the Cartesian coordinates of the vertices.

\begin{table}[!ht]
\begin{center}
\begin{tabular}{|c|c|c|c|}
\hline
vertex&$x$&$y$&$z$\\
\hline
A&$1/\sqrt{3}$&$1/\sqrt{3}$&$1/\sqrt{3}$\\
B&$-1/\sqrt{3}$&$1/\sqrt{3}$&$1/\sqrt{3}$\\
C&$-1/\sqrt{3}$&$-1/\sqrt{3}$&$1/\sqrt{3}$\\
D&$1/\sqrt{3}$&$-1/\sqrt{3}$&$1/\sqrt{3}$\\
\hline
E&$1/\sqrt{3}$&$1/\sqrt{3}$&$-1/\sqrt{3}$\\
F&$-1/\sqrt{3}$&$1/\sqrt{3}$&$-1/\sqrt{3}$\\
G&$-1/\sqrt{3}$&$-1/\sqrt{3}$&$-1/\sqrt{3}$\\
H&$1/\sqrt{3}$&$-1/\sqrt{3}$&$-1/\sqrt{3}$\\
\hline
\end{tabular}
\end{center}
\caption{Cartesian coordinates of the vertices of the cube.
Same conventions as in Table~\ref{ctetra}.}
\label{ccube}
\end{table}

\begin{figure}[!ht]
\begin{center}
\includegraphics[angle=90,width=.4\textwidth]{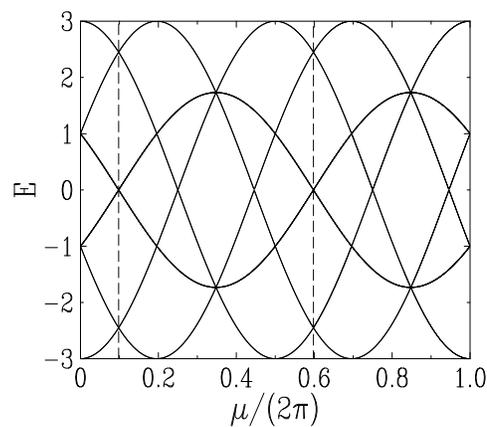}
\caption{
\label{spcube}
Plot of the energy spectrum of the cube
against $\mu/(2\pi)$ over one period.
Same conventions as in Figure~\ref{sptetra}.}
\end{center}
\end{figure}

The energy eigenvalues of the $16\times16$ Hamiltonian matrix
constructed from these coordinates
are listed in Table~\ref{tabcube} and shown in Figure~\ref{spcube}
as a function of $\mu/(2\pi)$ over one period.
The spectrum is observed to be its own opposite,
i.e., to be symmetric with respect to the origin of energies, $E=0$.
This extra symmetry is particular to the cube,
being due to the fact that this polyhedron is {\it bipartite}.

\begin{table}[!ht]
\begin{center}
\begin{tabular}{|c|c|c|}
\hline
$a$&$E_a(\mu)$&$m_a$\\
\hline
1&$3\cos\mu$&2\\
2&$\cos\mu+2\sqrt{2}\,\sin\mu$&2\\
\hline
3&$\cos\mu-\sqrt{2}\,\sin\mu$&4\\
4&$-\cos\mu+\sqrt{2}\,\sin\mu$&4\\
\hline
5&$-\cos\mu-2\sqrt{2}\,\sin\mu$&2\\
6&$-3\cos\mu$&2\\
\hline
\end{tabular}
\end{center}
\caption{Energy levels $E_a(\mu)$ of the cube
and their multiplicities $m_a$.
Same conventions as in Table~\ref{tabtetra}.}
\label{tabcube}
\end{table}

\subsection{The octahedron}
\label{specocta}

\begin{figure}[!ht]
\begin{center}
\includegraphics[angle=90,width=.35\textwidth]{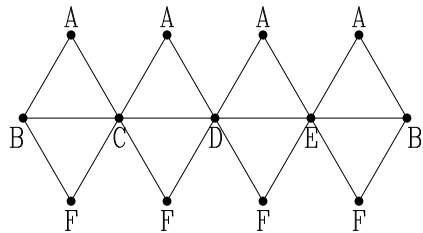}
\caption{
\label{vocta}
Planar representation of the octahedron.}
\end{center}
\end{figure}

The planar representation of the octahedron is shown in Figure~\ref{vocta}.
Table~\ref{cocta} lists the Cartesian coordinates of the vertices.

\begin{table}[!ht]
\begin{center}
\begin{tabular}{|c|c|c|c|}
\hline
vertex&$x$&$y$&$z$\\
\hline
A&0&0&1\\
\hline
B&1&0&0\\
C&0&1&0\\
D&$-1$&$0$&0\\
E&0&$-1$&0\\
\hline
F&0&0&$-1$\\
\hline
\end{tabular}
\end{center}
\caption{Cartesian coordinates of the vertices of the octahedron.
Same conventions as in Table~\ref{ctetra}.}
\label{cocta}
\end{table}

\begin{figure}[!ht]
\begin{center}
\includegraphics[angle=90,width=.4\textwidth]{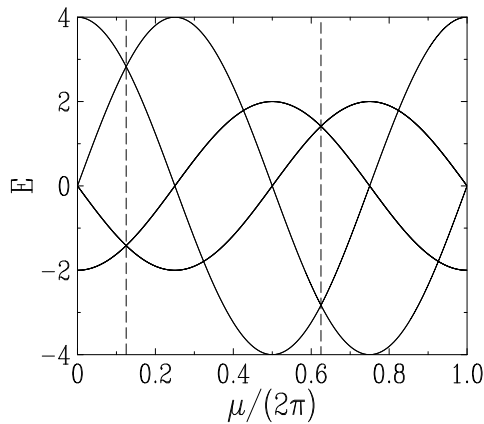}
\caption{
\label{spocta}
Plot of the energy spectrum of the octahedron
against $\mu/(2\pi)$ over one period.
Same conventions as in Figure~\ref{sptetra}.}
\end{center}
\end{figure}

The energy eigenvalues of the $12\times12$ Hamiltonian matrix
constructed from these coordinates
are listed in Table~\ref{tabocta} and shown in Figure~\ref{spocta}
as a function of $\mu/(2\pi)$ over one period.

\begin{table}[!ht]
\begin{center}
\begin{tabular}{|c|c|c|}
\hline
$a$&$E_a(\mu)$&$m_a$\\
\hline
1&$4\cos\mu$&2\\
2&$4\sin\mu$&2\\
\hline
3&$-2\sin\mu$&4\\
4&$-2\cos\mu$&4\\
\hline
\end{tabular}
\end{center}
\caption{Energy levels $E_a(\mu)$ of the octahedron
and their multiplicities $m_a$.
Same conventions as in Table~\ref{tabtetra}.}
\label{tabocta}
\end{table}

\subsection{The dodecahedron}
\label{specdode}

\begin{figure}[!ht]
\begin{center}
\includegraphics[angle=90,width=.4\textwidth]{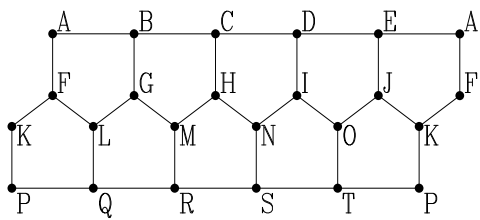}
\caption{
\label{vdode}
Planar representation of the dodecahedron.}
\end{center}
\end{figure}

The planar representation of the dodecahedron is shown in Figure~\ref{vdode}.
Table~\ref{cdode} lists the Cartesian coordinates of the vertices,
with the shorthand notations $c_k=\cos(k\pi/5)$, $s_k=\sin(k\pi/5)$, and
\beqa
&&a=\sqrt{\frac{2(5-\sqrt{5})}{15}},\quad
a'=\sqrt{\frac{5+2\sqrt{5}}{15}},\nonumber\\
&&b=\sqrt{\frac{2(5+\sqrt{5})}{15}},\quad
b'=\sqrt{\frac{5-2\sqrt{5}}{15}}.
\label{shortdode}
\eeqa

\begin{table}[!ht]
\begin{center}
\begin{tabular}{|c|c|c|c|}
\hline
vertex&$x$&$y$&$z$\\
\hline
A&$a\,c_0$&$a\,s_0$&$a'$\\
B&$a\,c_2$&$a\,s_2$&$a'$\\
C&$a\,c_4$&$a\,s_4$&$a'$\\
D&$a\,c_6$&$a\,s_6$&$a'$\\
E&$a\,c_8$&$a\,s_8$&$a'$\\
\hline
F&$b\,c_0$&$b\,s_0$&$b'$\\
G&$b\,c_2$&$b\,s_2$&$b'$\\
H&$b\,c_4$&$b\,s_4$&$b'$\\
I&$b\,c_6$&$b\,s_6$&$b'$\\
J&$b\,c_8$&$b\,s_8$&$b'$\\
\hline
\end{tabular}
\begin{tabular}{|c|c|c|c|}
\hline
vertex&$x$&$y$&$z$\\
\hline
K&$b\,c_9$&$b\,s_9$&$-b'$\\
L&$b\,c_1$&$b\,s_1$&$-b'$\\
M&$b\,c_3$&$b\,s_3$&$-b'$\\
N&$b\,c_5$&$b\,s_5$&$-b'$\\
O&$b\,c_7$&$b\,s_9$&$-b'$\\
\hline
P&$a\,c_9$&$a\,s_9$&$-a'$\\
Q&$a\,c_1$&$a\,s_1$&$-a'$\\
R&$a\,c_3$&$a\,s_3$&$-a'$\\
S&$a\,c_5$&$a\,s_5$&$-a'$\\
T&$a\,c_7$&$a\,s_7$&$-a'$\\
\hline
\end{tabular}
\end{center}
\caption{Cartesian coordinates of the vertices of the dodecahedron.
Same conventions as in Table~\ref{ctetra}.
Shorthand notations are explained in and above~(\ref{shortdode}).}
\label{cdode}
\end{table}

\begin{figure}[!ht]
\begin{center}
\includegraphics[angle=90,width=.45\textwidth]{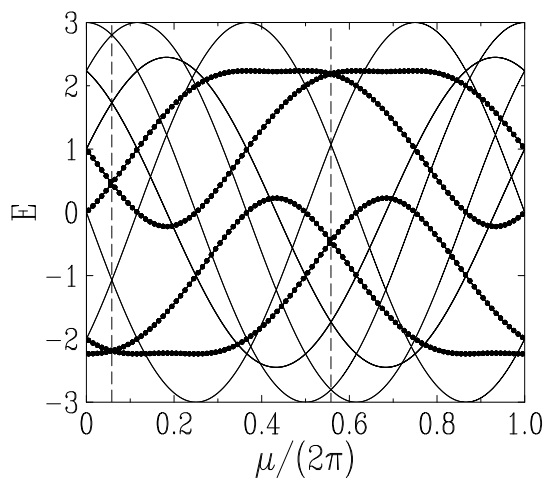}
\caption{
\label{spdode}
Plot of the energy spectrum of the dodecahedron
against $\mu/(2\pi)$ over one period.
Same conventions as in Figure~\ref{sptetra}.
Thick dotted lines:
sixfold degenerate energy levels $\eps_i(\mu)$ ($i=1,2,3,4$).}
\end{center}
\end{figure}

The energy eigenvalues of the $40\times40$ Hamiltonian matrix
constructed from these coordinates
are listed in Table~\ref{tabdode} and shown in Figure~\ref{spdode}
as a function of $\mu/(2\pi)$ over one period.
This is the first example where some of the energy levels are not
given by linear functions of $\cos\mu$ and $\sin\mu$.
The four sixfold degenerate energy levels $\eps_i(\mu)$ ($i=1,2,3,4$)
are the roots of the polynomial equation
\beq
\eps^4+2\xi\eps^3+(2\xi^2-5)\eps^2+2\xi(\xi^2-4)\eps+1-3\xi^2+\xi^4=0,
\label{po}
\eeq
with
\beq
\xi=\frac{\sqrt{5}+1}{2}\cos\mu+\frac{\sqrt{5}-1}{2}\sin\mu
=\sqrt{3}\,\cos(\mu-\Theta/2),
\eeq
and where the branches are chosen such that
$\eps_1(0)=1$, $\eps_2(0)=0$, $\eps_3(0)=-2$, $\eps_4(0)=-\sqrt{5}$.

\begin{table}[!ht]
\begin{center}
\begin{tabular}{|c|c|c|}
\hline
$a$&$E_a(\mu)$&$m_a$\\
\hline
1&$3\cos\mu$&2\\
2&$\sqrt{5}\,\cos\mu+2\sin\mu$&2\\
\hline
3&$\sqrt{5}\,\cos\mu-\sin\mu$&4\\
4&$\cos\mu+\sqrt{5}\,\sin\mu$&4\\
\hline
5&$\eps_1(\mu)$&6\\
6&$\eps_2(\mu)$&6\\
\hline
7&$-3\sin\mu$&2\\
8&$-2\cos\mu+\sqrt{5}\,\sin\mu$&2\\
\hline
9&$\eps_3(\mu)$&6\\
10&$\eps_4(\mu)$&6\\
\hline
\end{tabular}
\end{center}
\caption{Energy levels $E_a(\mu)$ of the dodecahedron
and their multiplicities $m_a$.
Same conventions as in Table~\ref{tabtetra}.
The functions $\eps_i(\mu)$ ($i=1,2,3,4$)
are the roots of the polynomial equation~(\ref{po}).}
\label{tabdode}
\end{table}

\subsection{The icosahedron}
\label{specico}

\begin{figure}[!ht]
\begin{center}
\includegraphics[angle=90,width=.4\textwidth]{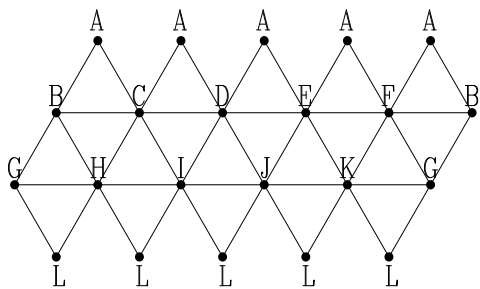}
\caption{
\label{vico}
Planar representation of the icosahedron.}
\end{center}
\end{figure}

The planar representation of the icosahedron is shown in Figure~\ref{vico}.
Table~\ref{cico} lists the Cartesian coordinates of the vertices,
with the shorthand notations $c_k=\cos(k\pi/5)$, $s_k=\sin(k\pi/5)$, and
\beq
d=\frac{2\sqrt{5}}{5},\quad d'=\frac{\sqrt{5}}{5}.
\label{shortico}
\eeq

\begin{table}[!ht]
\begin{center}
\begin{tabular}{|c|c|c|c|}
\hline
vertex&$x$&$y$&$z$\\
\hline
A&0&0&1\\
\hline
B&$d\,c_0$&$d\,s_0$&$d'$\\
C&$d\,c_2$&$d\,s_2$&$d'$\\
D&$d\,c_4$&$d\,s_4$&$d'$\\
E&$d\,c_6$&$d\,s_6$&$d'$\\
F&$d\,c_8$&$d\,s_8$&$d'$\\
\hline
\end{tabular}
\begin{tabular}{|c|c|c|c|}
\hline
vertex&$x$&$y$&$z$\\
\hline
G&$d\,c_9$&$d\,s_9$&$-d'$\\
H&$d\,c_1$&$d\,s_1$&$-d'$\\
I&$d\,c_3$&$d\,s_3$&$-d'$\\
J&$d\,c_5$&$d\,s_5$&$-d'$\\
K&$d\,c_7$&$d\,s_7$&$-d'$\\
\hline
L&0&0&$-1$\\
\hline
\end{tabular}
\end{center}
\caption{Cartesian coordinates of the vertices of the icosahedron.
Same conventions as in Table~\ref{ctetra}.
Shorthand notations are explained in and above~(\ref{shortico}).}
\label{cico}
\end{table}

\begin{figure}[!ht]
\begin{center}
\includegraphics[angle=90,width=.45\textwidth]{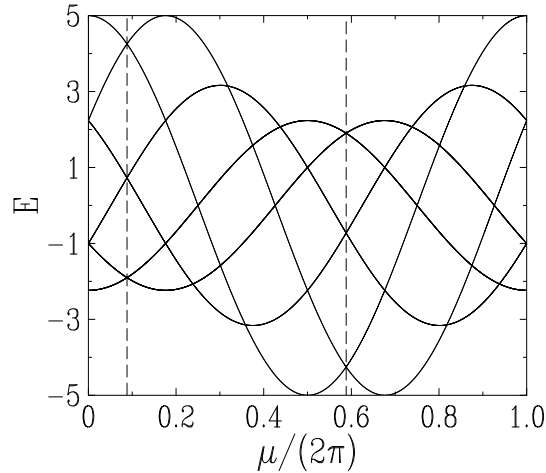}
\caption{
\label{spico}
Plot of the energy spectrum of the icosahedron
against $\mu/(2\pi)$ over one period.
Same conventions as in Figure~\ref{sptetra}.}
\end{center}
\end{figure}

The energy eigenvalues of the $24\times24$ Hamiltonian matrix
constructed from these coordinates
are listed in Table~\ref{tabico} and shown in Figure~\ref{spico}
as a function of $\mu/(2\pi)$ over one period.

\begin{table}[!ht]
\begin{center}
\begin{tabular}{|c|c|c|}
\hline
$a$&$E_a(\mu)$&$m_a$\\
\hline
1&$5\cos\mu$&2\\
2&$\sqrt{5}(\cos\mu+2\sin\mu)$&2\\
\hline
3&$\sqrt{5}(\cos\mu-\sin\mu)$&4\\
4&$-\cos\mu+3\sin\mu$&4\\
\hline
5&$-\cos\mu-2\sin\mu$&6\\
6&$-\sqrt{5}\,\cos\mu$&6\\
\hline
\end{tabular}
\end{center}
\caption{Energy levels $E_a(\mu)$ of the icosahedron
and their multiplicities $m_a$.
Same conventions as in Table~\ref{tabtetra}.}
\label{tabico}
\end{table}

\subsection{The fullerene}
\label{specfulle}

We now turn to the case of the C$_{60}$ fullerene.
For simplicity we model this molecule as a symmetric truncated icosahedron,
where all the links have equal lengths,
so that the analysis of Section~\ref{defs} applies.
Let us however recall
that this symmetry is known to be slightly violated~\cite{lengths},
as for the free molecule
the length of the sides of the pentagons is 1.46 \AA,
whereas the length of the other links is 1.40 \AA.

\begin{figure}[!ht]
\begin{center}
\includegraphics[angle=90,width=.7\textwidth]{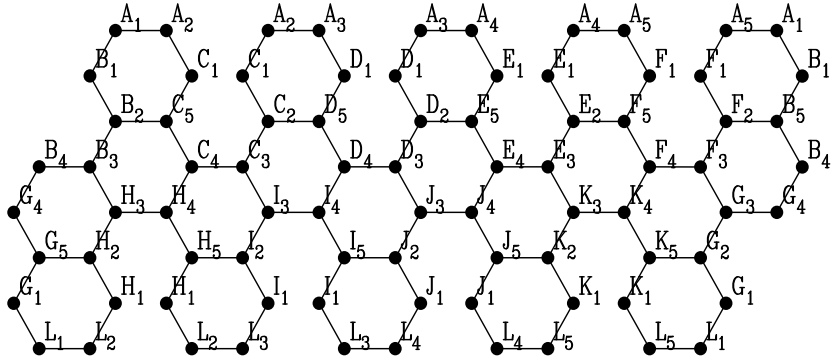}
\caption{
\label{vfulle}
Planar representation of the fullerene (symmetric truncated icosahedron).}
\end{center}
\end{figure}

The symmetric truncated icosahedron has $V=60$ equivalent vertices,
$L=90$ equivalent links, and $F=32$ faces, namely 12 pentagons and 20 hexagons,
respectively corresponding to the vertices and to the faces of the icosahedron.
Figure~\ref{vfulle} shows the planar representation
obtained by unwrapping the fullerene around a fivefold axis
going through the opposite pentagonal faces
${\rm A}_1\dots{\rm A}_5$ and ${\rm L}_1\dots{\rm L}_5$.

\begin{figure}[!ht]
\begin{center}
\includegraphics[angle=90,width=.2\textwidth]{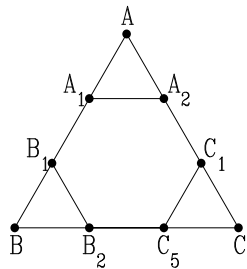}
\caption{
\label{tri}
The triangular face ABC of the icosahedron
decorated by vertices of the fullerene.
Notations are consistent with Figures~\ref{vico} and~\ref{vfulle}.}
\end{center}
\end{figure}

The Cartesian coordinates of the vertices of the fullerene have been derived
from those of the vertices of the icosahedron, listed in Table~\ref{cico},
using the approach described in the Appendix of~\cite{I}.
This procedure is illustrated in Figure~\ref{tri},
showing an enlargement of the upper left part
of Figures~\ref{vico} and~\ref{vfulle}, with consistent notations.
One has
\beq
\ve A_1=\lambda(2\ve A+\ve B),\quad\ve A_2=\lambda(2\ve A+\ve C),
\eeq
and so on, with
\beq
\lambda=\sqrt\frac{25-4\sqrt{5}}{109},
\eeq
so that
\beq
\cos\Theta=\ve A_1\cdot\ve A_2=(4+\sqrt{5})\lambda^2=\frac{80+9\sqrt{5}}{109}.
\label{thfulle}
\eeq

\begin{figure}[!ht]
\begin{center}
\includegraphics[angle=90,width=.55\textwidth]{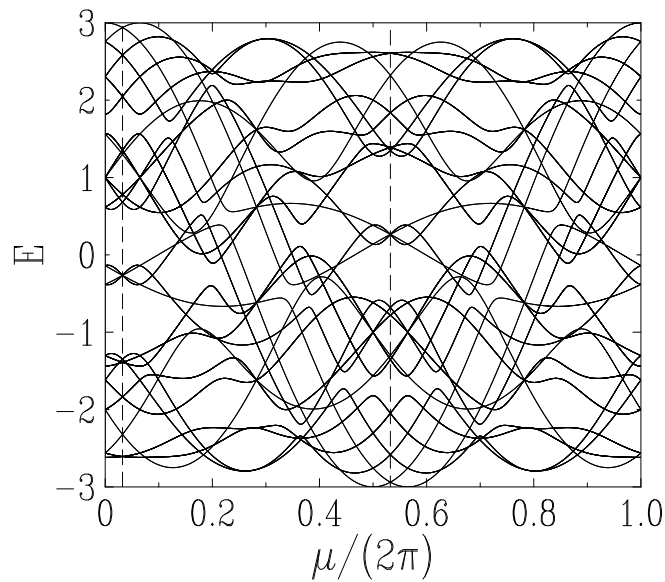}
\caption{
\label{spfulle}
Plot of the energy spectrum of the fullerene
against $\mu/(2\pi)$ over one period.
Same conventions as in Figure~\ref{sptetra}.}
\end{center}
\end{figure}

The energy eigenvalues of the $120\times120$ Hamiltonian matrix
constructed from the coordinates thus obtained have been evaluated
by means of a numerical diagonalization.
The energy spectrum is shown in Figure~\ref{spfulle}
as a function of $\mu/(2\pi)$ over one period.

For $\mu=0$, i.e., in the absence of spin-orbit coupling,
we recover two independent copies of
the known tight-binding spectrum of the fullerene~\cite{manou},
with its 15 distinct energy levels with multiplicities ranging from 1 to 9.
For generic non-zero values of $\mu$,
the spectrum consists of 28 distinct energy levels
with multiplicities ranging from 2 to~6 only.
As $\mu\to0$ the 28 levels merge into the 15 ones
according to the patterns given in Table~\ref{tabfulle}.
We have introduced the shorthand notation
\beq
w_\pm=\sqrt{2(19\pm\sqrt{5})}.
\label{wpm}
\eeq

\begin{table}[!ht]
\begin{center}
\begin{tabular}{|c|c|c|c|}
\hline
$E(0)$&$E(0)_{\rm num}$&$m(0)$&$m(\mu\to0)$\\
\hline
$3$&$3$&$2$&$2$\\
$(3+\sqrt{5}+w_-)/4$&$2.756598$&$6$&$2+4$\\
$(\sqrt{13}+1)/2$&$2.302776$&$10$&$4+6$\\
$(3-\sqrt{5}+w_+)/4$&$1.820249$&$6$&$6$\\
$(\sqrt{17}-1)/2$&$1.561553$&$8$&$6+2$\\
$1$&$1$&$18$&$6+2+6+4$\\
$(\sqrt{5}-1)/2$&$0.618034$&$10$&$6+4$\\
$(3+\sqrt{5}-w_-)/4$&$-0.138564$&$6$&$4+2$\\
$(\sqrt{5}-3)/2$&$-0.381966$&$6$&$2+4$\\
$-(\sqrt{13}-1)/2$&$-1.302776$&$10$&$4+6$\\
$(3-\sqrt{5}-w_+)/4$&$-1.438283$&$6$&$6$\\
$-(\sqrt{5}+1)/2$&$-1.618034$&$10$&$4+6$\\
$-2$&$-2$&$8$&$6+2$\\
$-(\sqrt{17}+1)/2$&$-2.561553$&$8$&$2+6$\\
$-(\sqrt{5}+3)/2$&$-2.618034$&$6$&$6$\\
\hline
\end{tabular}
\end{center}
\caption{Energy levels $E(0)$ of the fullerene at $\mu=0$,
with their multiplicities $m(0)$
and degeneracy-lifting patterns at small $\mu\ne0$.
The shorthand notation $w_\pm$ has been introduced in~(\ref{wpm}).}
\label{tabfulle}
\end{table}

\section{Total energy}
\label{TotalE}

An interesting illustration of the above energy spectra is provided
by the total energy at half filling, defined as
\beq
\E=\sum_{a=1}^{V} E_a,
\label{etotdef}
\eeq
where the $2V$ energy levels are assumed to be in increasing order
($E_1\le E_2\le\dots\le E_{2V}$) and repeated according to their multiplicities.

The first of the sum rules~(\ref{sumrules})
implies that the total energy thus defined is insensitive to the
sign of the Hamiltonian $\hat\H$.
Combining this feature with the symmetries derived
in Section~\ref{props}, we conclude that $\E(\mu)$ obeys the symmetries
\beq
\E(\mu)=\E(\Theta-\mu)=\E(\mu+\pi).
\eeq
The total energy therefore has period $\pi$,
and exhibits two inequivalent stationary points per period, at
\beq
\mu_0=\Theta/2,\quad\mu_{\rm m}=(\Theta+\pi)/2.
\label{stats}
\eeq
The first of these values, $\mu_0$, coincides with one of the
special values introduced in~(\ref{spec}), i.e., one of
the symmetry axes of the spectrum.
The second of the above values, $\mu_{\rm m}$, corresponds to one of the
midpoints between the latter symmetry axes.

The second of the sum rules~(\ref{sumrules})
implies that the mean squared value of the individual energy levels is
$\mean{E^2}=4L/(2V)=p$, where $p$ is the coordination number of the vertices.
This suggests to introduce the reduced total energy
\beq
\E_\rr=\frac{\E}{2V\!\sqrt{\mean{E^2}}}
=\frac{\E}{2V\!\sqrt{p}}=\frac{\E}{2\sqrt{2VL}}.
\label{erdef}
\eeq
This heuristic argument can be turned to a quantitative prediction
in the $p\to\infty$ limit of a very highly connected structure~\cite{I}.
In this limit, the reduced total energy $\E_\rr$
has been shown to have the universal limiting value
\beq
\E_\infty=-\frac{1}{\sqrt{2\pi}}=-0.398942.
\label{elimit}
\eeq

\begin{figure}[!ht]
\begin{center}
\includegraphics[angle=90,width=.5\textwidth]{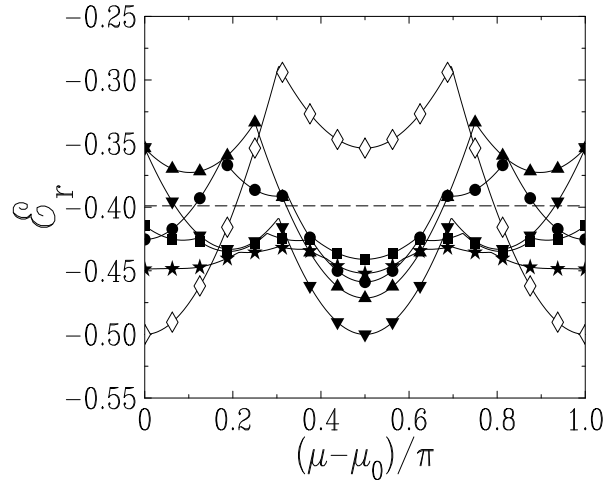}
\caption{
\label{etot}
Plot of the reduced total energy $\E_\rr$ against $(\mu-\mu_0)/\pi$,
for all the polyhedra investigated in this work:
tetrahedron ($V=4$) (empty diamonds), cube ($V=8$) (down triangles),
octahedron ($V=6$) (up triangles), dodecahedron ($V=20$) (squares),
icosahedron ($V=12$) (circles) and fullerene ($V=60$) (stars).
The curves consist of many more points than symbols
(500 data points for each polyhedron).
The horizontal dashed line shows the limiting value~(\ref{elimit}).}
\end{center}
\end{figure}

\begin{figure}[!ht]
\begin{center}
\includegraphics[angle=90,width=.5\textwidth]{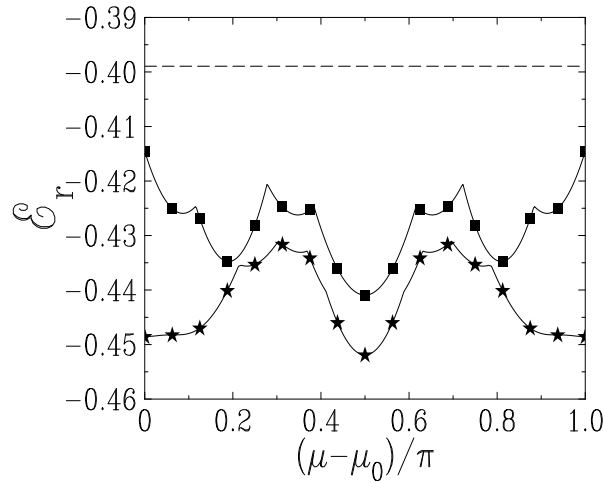}
\caption{
\label{etotdetail}
Enlargement of Figure~\ref{etot}, with the same conventions,
emphasizing the weak $\mu$-dependence of the reduced total energy
$\E_\rr$ in the cases of the dodecahedron and of the fullerene.}
\end{center}
\end{figure}

Figure~\ref{etot} shows a plot of the reduced total energy $\E_\rr$
for all the polyhedra investigated in this work,
as a function of $(\mu-\mu_0)/\pi$ over one period.
The reduced total energy is observed to wander
around the limiting value~(\ref{elimit}), shown as a dashed line.
The amplitude of the oscillations, i.e.,
of the dependence of the total energy on the parameter~$\mu$,
is a decreasing function of the number of vertices.
Figure~\ref{etotdetail} shows an enlargement of the plot
focusing on the weak $\mu$-dependence of $\E_\rr$
in the two examples with the larger numbers of vertices,
i.e., the dodecahedron ($V=20$) and the fullerene ($V=60$).
The abscissa axes in Figures~\ref{etot} and~\ref{etotdetail} are such that
the stationary point~$\mu_0$ introduced in~(\ref{stats})
corresponds to the ends of the plots,
whereas $\mu_{\rm m}$ corresponds to their centers.
The latter stationary point is observed to be
the absolute minimum of the total energy
for all the polyhedra considered in this work, except the tetrahedron,
for which the total energy has its absolute minimum
at $\mu=\mu_0$ and a local minimum at~$\mu=\mu_{\rm m}$.

\section{Discussion}
\label{discussion}

In this paper we have introduced and investigated a tight-binding model
defined on graphs drawn on the unit sphere,
describing the motion of an electron subject to a spin-orbit interaction
in the radial electric field created by a classical charge sitting
at the center of the sphere.
The present work completes our study of electronic properties
of mesoscopic and nanoscopic systems with the topology of the sphere,
started in the companion work~\cite{I}, which is devoted to electrons subject
to a radial magnetic field produced by a quantized
magnetic charge sitting at the center of the sphere.

This work has been focused onto polyhedral graphs such that all links
have a common arc length $\Theta$.
For a fixed graph of this kind, the model has only one parameter,~$\mu$,
giving a dimensionless measure of the strength of the spin-orbit inter\-action.
Among the symmetry properties of the model,
exposed in detail in Section~\ref{props},
the $\mu\leftrightarrow\Theta-\mu$ symmetry
was quite unexpected, as it has no counterpart in the continuum,
described by the familiar form $\ve L\cdot\ve S$ of the spin-orbit Hamiltonian,
whose eigenvalues and multiplicities are recalled in~(\ref{specLS}).

For the special value $\mu=\mu_0=\Theta/2$,
which coincides with one of the symmetry axes of the spectra,
an exact correspondence has been established with the tight-binding problem
in the magnetic field of a Dirac monopole,
investigated in~\cite{I}.
It is remarkable that, by tuning a parameter
in a theory which is experimentally realizable,
it is possible to obtain the spectrum of another system
whose experimental realization is so far elusive.
In fact this correspondence already holds at the classical level.
It was indeed discovered long ago by Poincar\'e~\cite{poincare}
that the motion of an electrically charged particle
in the field of a magnetic charge can be mapped onto that of a spherical top.
In a quantum-mechanical framework,
the quantitative correspondence between both problems reads~\cite{haldane,shnir}
\beq
\abs{n}=2S,
\label{smc}
\eeq
where the integer $n$ is the magnetic charge
(in units of the elementary magnetic charge of Dirac's monopole),
whereas $S$ is the total spin of the top.
The above relation can easily be recovered by noticing that the ground state
of the Schr\"odinger equation
on the sphere in the presence of a magnetic charge $n$,
investigated in the pioneering work of Tamm~\cite{tamm},
has a multiplicity $\abs{n}+1$, to be identified with $2S+1$.
In the present situation of an electron ($S=1/2$),
the correspondence has indeed been shown
to hold for a unit magnetic charge ($n=\pm1$).
More generally, a similar correspondence can be expected to hold true
for higher representations as well,
whenever the spin and the magnetic charge are related by~(\ref{smc}),
for a suitably chosen special form of the spin-orbit interaction.

We have then turned to the study of specific examples of polyhedra,
namely the five Platonic solids
(tetrahedron, cube, octahedron, dodecahedron and icosahedron)
and the C$_{60}$ fullerene (modeled as a regular truncated icosahedron).
For the Platonic solids, the full $\mu$-dependence of the energy levels,
and the corresponding multiplicities, have been obtained analytically
in Sections~\ref{spectetra}--\ref{specico}.
These results allow for an explicit check of the general
properties listed in Section~\ref{props}.
Rather surprisingly, all the energy levels can be expressed
as linear combinations of $\sin\mu$ and $\cos\mu$,
except the four sixfold degenerate levels of the dodecahedron,
which are obtained as the roots of a fourth-degree polynomial
given in~(\ref{po}).
This simplicity of the energy eigenvalues
is to be contrasted with the rather large dimension, $2V$,
of the Hamiltonian matrices, i.e., 40 for the dodecahedron.

Pursuing along the lines of our companion work~\cite{I},
we have also evaluated the total electronic energy $\E$
of the system at half filling.
For all the examples considered in this work,
this total energy is found to be rather close to its asymptotic value
in the limit of large coordination numbers,
where the density of states becomes Gaussian.
Finally, as far as its dependence on the parameter $\mu$ is concerned,
the total energy reaches its absolute minima at
the midpoints between the symmetry axes of the spectra
for all the polyhedra considered in this work, except for the tetrahedron
where the total energy has its absolute minima
at the symmetry axes of the spectrum and local minima at the midpoints.

\subsection*{Acknowledgments}

It is a pleasure for us to thank B.~Dou\c cot, G.~Montambaux
and O.~Entin-Wohlman for very stimulating discussions.

\end{document}